\documentclass[12pt,a4paper,final]{iopart}

\usepackage{iopams}  
\usepackage{graphicx}
\usepackage[breaklinks=true,colorlinks=true,linkcolor=blue,urlcolor=blue,citecolor=blue]{hyperref}

\expandafter\let\csname equation*\endcsname\relax
\expandafter\let\csname endequation*\endcsname\relax
\usepackage{amsmath}
\usepackage{mathtools}
\usepackage{doi}

\begin{document}
	\title[Embedding Human Heuristics in Machine-Learning-Enabled Probe Microscopy]{Embedding Human Heuristics in Machine-Learning-Enabled Probe Microscopy}
	
	\author[cor1]{Oliver M. Gordon}
	\address{School of Physics \& Astronomy, The University of Nottingham, University Park, Nottingham, NG7 2RD,
		United Kingdom}
	\ead{oliver.gordon@nottingham.ac.uk}
	
	\author{Filipe L.Q. Junqueira}
	\address{School of Physics \& Astronomy, The University of Nottingham, University Park, Nottingham, NG7 2RD,
		United Kingdom}
	\ead{filipe.junqueira@nottingham.ac.uk}
	
	\author{Philip J. Moriarty}
	\address{School of Physics \& Astronomy, The University of Nottingham, University Park, Nottingham, NG7 2RD,
			United Kingdom}
	\ead{\mailto{philip.moriarty@nottingham.ac.uk}}

\begin{abstract}
	Scanning probe microscopists generally do not rely on complete images to assess the quality of data acquired during a scan. Instead, assessments of the state of the tip apex, which not only determines the resolution in any scanning probe technique but can also generate a wide array of frustrating artefacts, are carried out in real time on the basis of a few lines of an image (and, typically, their associated line profiles.) The very small number of machine learning approaches to probe microscopy published to date, however, involve classifications based on full images. Given that data acquisition is the most time-consuming task during routine tip conditioning, automated methods are thus currently extremely slow in comparison to the tried-and-trusted strategies and heuristics used routinely by probe microscopists. Here, we explore various strategies by which different STM image classes (arising from changes in the tip state) can be correctly identified from partial scans. By employing a secondary temporal network and a rolling window of a small group of individual scanlines, we find that tip assessment is possible with a small fraction of a complete image. We achieve this with little-to-no performance penalty -- or, indeed, markedly improved performance in some cases -- and introduce a protocol to detect the state of the tip apex in real time. 
\end{abstract}

\vspace{2pc}
\submitto{\MLST}
\newpage
\section{Introduction}

	One of the major challenges in the drive to fully automate the scanning probe microscope is the need to constantly maintain the integrity of the tip\cite{Tajaddodianfar2018,Tewari2017}. During an experimental session, interactions with the surface can cause the tip to spontaneously and randomly change shape, modifying the interactions and therefore changing the data acquired in a highly non-linear fashion. This frequently results in inconsistent scans containing visual artefacts, often making data unusable or, at best, problematic to interpret. Furthermore, it is becoming \textit{de rigeuer} in state-of-the-art SPM to functionalise tips by deliberately picking up adsorbed molecules or atoms from the surface\cite{Giessibl2019}, vastly improving resolution\cite{Gross2009}, enabling direct measurement of intermolecular pair potentials\cite{Sun2011,Chiutu2012}, and/or modifying the capability of the probe, for better or worse, to manipulate and position single adsorbates\cite{Meyer2001}. \\
	
	Indeed, SPM experimentation is now at the point where not only is single atom/molecule termination of the tip apex required, but fine control and detailed understanding of its atomic/molecular orbital structure is often essential. Gross \emph{et al.}\cite{Gross2011} provided a particularly elegant example of the importance of ``orbital engineering'' of this type by demonstrating the significant enhancement of submolecular resolution in scanning tunnelling microscopy (STM) images of pentacene and naphthalocyanine molecules via tunnelling through $p$-wave orbitals, as the tunnelling matrix element for these states is proportional not to the sample wavefunction itself but its spatial derivatives. The spatial distribution and orientation of electron density at the tip apex also plays a central role in single atom manipulation\cite{Jarvis2012}. Controlling and maintaining the atomistic and orbital structure of the tip apex is therefore a vital part of state-of-the-art SPM operation. Currently, this requires a protracted and repetitive routine of voltage pulsing, ``gentle'' (or not-so-gentle) indenting of the tip into the surface, scanning at relatively high voltages and currents, and/or attempts to pick up adsorbates. This is at present a high-effort, time-consuming and manual process involving only simple sub-processes, making it ideal to automate. \\
	
	Whilst convolutional neural networks (CNNs) have been shown to be capable of assessing SPM tips\cite{Rashidi2018, Rashidi2019,Gordon2019}, and, most recently, of extracting ``hidden order'' from STM datasets\cite{Zhang2019}, CNN methods to-date have been trained exclusively with complete images. Partial scans comprising a small number of scanlines therefore simply do not provide the information upon which the network mathematically depends and so current methods of CNN image assessment require complete scans. This method of CNN assessment \textit{after complete} scans compares extremely poorly to human-based assessment, in which SPM operators routinely perform accurate assessment \textit{during in-process} scans by observing individual line profiles as the image is acquired. Indeed, as little as 1-2\% of a full scan may be required to correctly assess a particularly poor tip. Furthermore, because the majority of time spent maintaining the tip is spent acquiring the data to assess, manual maintenance by a human is beyond an order-of-magnitude faster than any current CNN protocol. Given that manual maintenance can take several hours as-is, automated tip assessment with full-scan CNN protocols may be unable to keep up with the demands of SPM experimentalists unless an alternative strategy is introduced. In this paper we outline such a strategy and demonstrate that it performs extremely well against current methods based on complete images.  \\
	
	Beginning with a dataset of 6167 scans of the H:Si(100) surface, we extend our previous study of CNN tip detection\cite{Gordon2019} to explore and compare a variety of methods by which the state of the tip can be determined using incomplete, partial scans. In addition to the simple, common method of "padding" incomplete scan frames with an arbitrary marker value, we also discuss training the network to recognise individual linescans instead of entire images. Optimal performance is seen when classifying a "window" consisting of a small group of linescans, and using a second temporal network to determine tip state as the window is "rolled" over the course of a scan. This method remarkably produces better-than-complete-image performance with only a fraction of the data. By combining several of these methods in a "hybrid" approach, it is possible to accurately assess scanning probe (in this case, STM) data by at least an order of magnitude faster than current CNN protocols\cite{Rashidi2018, Rashidi2019, Gordon2019}.

\section{H:Si(100) Dataset}

\begin{figure}[htb!]
		\centering
		\includegraphics[width=\textwidth]{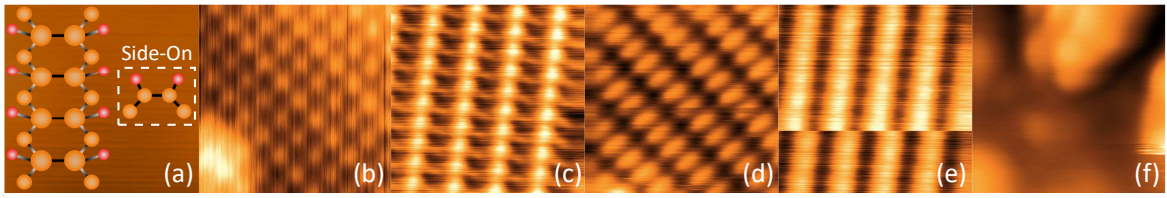}
		\caption{Selection of images showing key tip states for STM imaging of H:Si(100). (a) Ball-and-stick model of atomic structure of H:Si(100)-(2x1), which comprises rows of hydrogen-terminated silicon dimers; (b) atomic resolution; (c) asymmetry; (d) dimer resolution; (e) row resolution; and (f) bad/blurry. The tip can also spontaneously change during imaging, resulting in the horizontal discontinuity in (e). Frequently, features appear to blend between images, such as with asymmetries and atoms, or the dimer-like modulation in rows. Asymmetries and dimer classes were therefore combined.}
		\label{fig:tip_states}
	\end{figure}

	As discussed in the Introduction, SPM images often contain multiple features because of the scanning probe apex changing during a single scan. These tip changes also regularly and immediately result in discontinuities perpendicular to the direction of the scan. After the tip changes shape, multiple, more complex visual artefacts can also appear\cite{Straton2014,Woolley2011,Stirling2013}. For example, features can appear to 'ghost' due to the presence of multiple tip apices\cite{Wang2016,Rashidi2018}, or large blurs may appear due to impurities on the probe itself. Whilst these particular features can be seen when scanning any surface, others are specific to the surface being investigated\cite{Wolkow1992}. For example, for the H:Si(100) surface, four different, distinct tip states of 'individual atoms' (for the sharpest tips), 'dimers, 'asymmetries', and 'rows' have been observed and discussed in the literature\cite{Woolley2011,Sweetman,Sweetman2012}. Typically, an operator will want to coerce the tip into producing images with one of these atomistic resolutions visible. (It is also worth noting that the tip apex capable of the highest resolution may not be best suited to other tasks, including, in particular, single atom manipulation\cite{Moller2017}.) Uncontrolled, and sometimes controlled, tip changes, however, mean that it is possible to produce images of H:Si(100) showing a combination of any of these four states, tip change shears, and other defects. Examples for each state are shown in Figure \ref{fig:tip_states}, along with a diagram of the H:Si(100)-(2x1) surface reconstruction. \\
	
	\newpage
	Besides its distinctive surface features, H:Si(100) is an ideal test-bed for developing CNN automation techniques. In addition to the relative simplicity of its reconstruction and a wealth of previous literature\cite{WalshHersam2009}, H:Si(100) is a well understood substrate that has been used in many important advances in single atom technology and atomically precise materials engineering\cite{Shen1995,Lopinski2000,Fuechsle2012, Weber2012,Moller2017,Achal2017,Huff2018}. Furthermore, because it has been previously studied in the context of machine-learning-enabled SPM\cite{Rashidi2018, Rashidi2019, Gordon2019}], a good comparison can be formed with existing machine learning approaches based on full scans. As such, we used our existing dataset of 6167 complete images of H:Si(100)\cite{Gordon2019}. These images were acquired on a Omicron variable-temperature STM between March 2014 and November 2015, and at varying scan sizes and voltage biases of 3x3nm\textsuperscript{2} to 80x80nm\textsuperscript{2} and -2V to +2V respectively. They were then hand-classified into the four categories listed above, as well as "tip changes" and "generic defects". Specific defects were not considered, as tip conditioning is performed based on the presence of any defect, and not the specific defect itself. As such, combining all defects into one category simplified the classification task, improving CNN performance. \\
	
	From here, images were then randomly assigned into a training/testing set for training the network. Performance was then calculated with a separate, blind holdout set for verification. After random shuffling, 4987 of the images were assigned into the training/testing datasets in an 80/20 split, and 1180 into the holdout set. Data was then filtered to remove ambiguous images that were classified in multiple categories and/or the human classifiers did not perfectly agree upon\cite{Gordon2019}. This left 3386 images for training/testing, and 648 for blind verification. Because of the relatively small number of images available and to further improve performance, all data were then pre-processed using identical methods as in Gordon \emph{et al.}\cite{Gordon2019} (namely flattening and scaling linescans to have mean of 0 and standard deviation of 1). The training/testing sets were also augmented with vertical and horizontal flips, random rotations from 0-360$^{\circ}$, crops, pans, and random Gaussian noise. This step was needed to prevent the network from rapidly overfitting. (This is where a CNN learns about random noise in the training set\cite{goodfellow2016deep}, performing extremely well during training, but poorly with testing/verification data unseen during training.) To allow for the best approximation of real-world performance on unseen data, the verification set was not augmented.\\
	
	Furthermore, the data were also downscaled, which reduced both training time and overfitting further still\cite{goodfellow2016deep}. Previously, optimal performance was found when reducing full-scans from size 512x512 to 128x128\cite{Gordon2019}. Panning and cropping augmentations were applied in such a way that 128x128 regions of the images were taken, allowing for downscaling without interpolation of data. Because the holdout data were not augmented, these images were downscaled in the more traditional sense.	 \\

	Because an operator may desire the presence of some tip states (e.g. 'individual atoms'), but desire the absence of others (e.g. ``blurry''/defects), there are different implications to predicting the tip to be (or not to be) in different states. This makes use of the fact that CNNs do not make binary yes/no predictions, but instead output confidence ratings between 0 and 1 for each category. A decision is then made by rounding each number to 0 or 1. This rounding can be altered, and true positive/false positive rates then compared to demonstrate the overall performance of the classifier for each category. This forms the receiver-operator characteristic curve (ROC)\cite{Fawcett2006, scikit-learn}, which is then easily quantified by calculating the area under (AU) the curve. The AUROC for all categories can then be averaged to give an average AUROC for the classifier as a whole. A perfect classifier has AUROC = 1, while a network that operates purely by guessing has AUROC = 0.5. This is independent of the number of images in each class, which drastically skews a pure accuracy value. As a result of this class imbalance, we therefore also calculate weighted accuracy\cite{scikit-learn} instead of pure accuracy.
	
\section{Results and Discussion}

	\subsection{Data Padding/Masking\label{sect:pad_mask}}
\begin{figure}[htb!]
			\centering
			\includegraphics{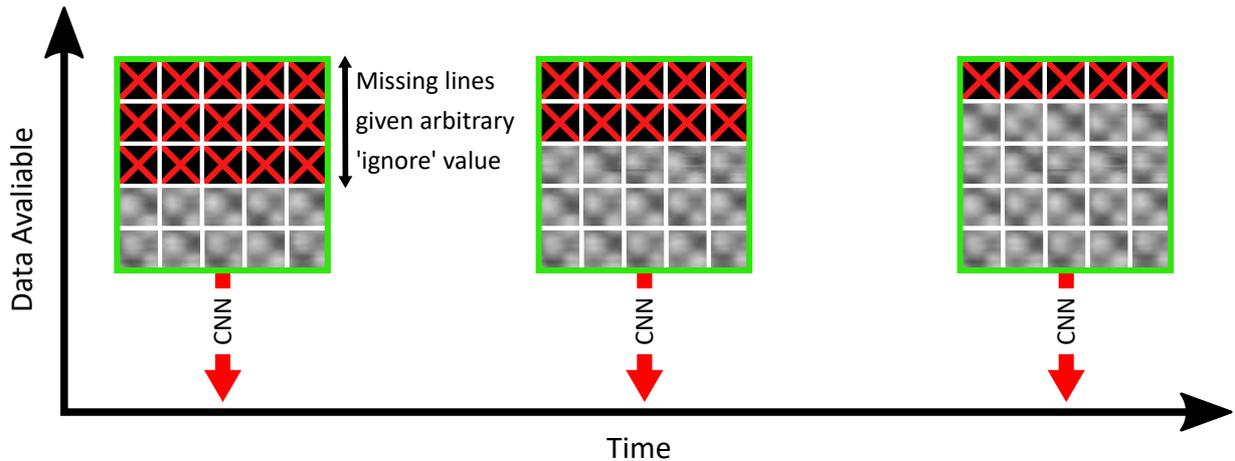}
			\caption{Figure to demonstrate a potential method to allow neural networks to predict the state of an SPM tip using incomplete scans. Because CNNs can only make predictions if given the same number of data points used during training, it is not possible to make predictions using incomplete scans. It is also computationally wasteful to create multiple CNNs for each stage of scan completeness. Instead, partial scans can be "padded" with an arbitrary marker value until there are enough data points to equal a full sized scan. After each successive linescan, less padding is required. This allows the CNN (green border) to train/predict using incomplete scans.}
			\label{fig:padding_method}
		\end{figure}

		In many neural network applications, data are often of varying length. For example, in natural language processing\cite{Young2018}, some words and sentences are inevitably longer than others. In these cases, shorter pieces of data are lengthened by "padding" them with a marker value\cite{Young2018} until they are as long as the longest piece of data. The marker value is chosen such that it cannot naturally appear in the real data. Training and testing then continues as normal, as the network learns to ignore the marker value. In the context of SPM, we can exploit the fact that images are sequentially generated one linescan at a time, and that completed images contain the same number of linescans, regardless of scan parameters. During an incomplete scan, the missing linescans can therefore be replaced with a marker value to allow the network to produce an output. Figure \ref{fig:padding_method} demonstrates how data could be padded during scanning to form a full sized image. \\
		
		As such, it is possible simulate and test partial scans with the original dataset of complete scans. To do this, a random number of linescans from the end of the scan were "masked" during training by replacing the real data with the marker value. To do this, we let		\\
		\begin{equation}\label{eq:padding}
			\begin{bmatrix*}[l]
				\mathbf{A}^i_j\\[0.3em]
				\mathbf{A}^i_{j+1}\\[0.3em]
				\mathbf{A}^i_{j+2}\\[0.3em]
				\mathbf{\hspace{0.9em}\vdots}\\[0.3em]
				\mathbf{A}^i_N
			\end{bmatrix*}
			= M,
		\end{equation}				
		where $N$ is the total number of lines in a full image, and $M$ the marker value. This produces an array, $\mathbf{A}^i_j$, for the $i^{th}$ image of a dataset, in which only $j$ linescans appear to have been produced. To improve performance, data is further augmented by repeating $\mathbf{A}^i$ multiple times, but with randomly assigned $j$.\\
						
		Although this method is simple and can easily be applied to existing protocols, the use of a marker value is of course highly problematic. In the context of SPM, data can theoretically contain any positive or negative value within the operating range of the acquisition hardware. As such, no marker value exists that could not show up in the actual dataset, without being so large as to make the actual data miniscule by comparison and negatively impacting learning. As such, the network will likely become insensitive to some of the actual data. Given that each line was pre-processed to have mean of 0 and standard deviation of 1, we therefore consider arbitrary marker values of $M=0$ and $M=10$. As an alternative, we also consider "tiling" by repeating $\mathbf{A}^i_j$ to full scan-size. This avoids the need to fill with an arbitrary marker value.\\
		
		The CNN structure was chosen to be VGG-like\cite{Simonyan2014} after strong all-round performance was previously found for H:Si(100) using a similar structure\cite{Gordon2019}. This network\cite{Simonyan2014} begins with two 2D convolutional layers of 32 output filters, 3x3 convolutional filters, and 3x3 strides. This is followed by a third max pooling layer with 2x2 convolutional filters and 2x2 strides. This three layer block is then repeated, but with output filters of 64 and then 128 layers respectively. The very first convolutional layer in the model was then altered to have 7x7 convolutional filters and 2x2 strides. This structure was then trained three separate times to create a majority voting ensemble. Not only does this allow for the performance benefits seen when taking a majority vote of a subjective task, but also reduces variance in CNN performance which was found to vary by about 1\% between repeats. \\
			
\begin{figure}[htb!]
			\centering
			\includegraphics{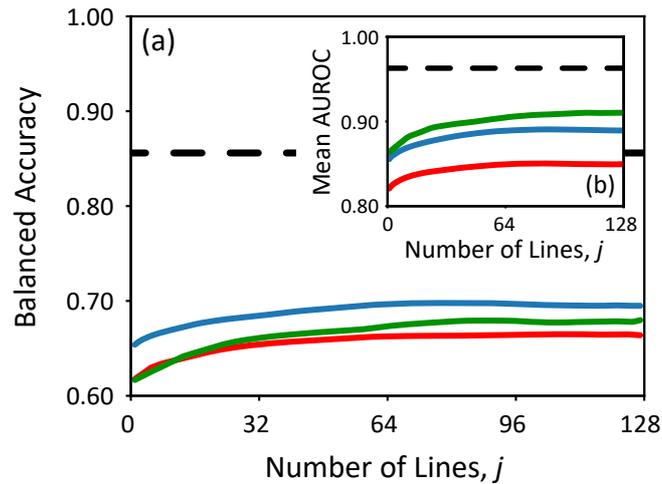}
			\caption{Figure to demonstrate the balanced accuracy (a), and mean area-under the receiver-operator-characteristic curve (b) of a neural network trained to classify partial STM images of the H:Si(100) surface. Given that SPM data is generated one line at a time, incomplete scans can be padded to full-size with a marker value that the otherwise identical network then learns to ignore. In this way, the data requirements for neural network automated state detection can be reduced significantly. Here, we consider marker values of 10 (red), 0 (blue), and also tile the data to size (green). However, performance is far below an identical network trained exclusively on full size data (black)}
			\label{fig:padding_performance_128}
		\end{figure}

		To test this method, the performance of the CNN ensemble was calculated as one additional line was unmasked at a time. We do this by masking from the $j^{th}$ line of $\mathbf{A}^i_j$ using Equation \ref{eq:padding} for all 648 images in the verification dataset. The CNN ensemble was then used to predict the tip state, $\mathbf{P}(\mathbf{A}^i_j)$, from $j=2$ to $j=N$. By assuming the human prediction to be perfectly correct, performance was calculated by comparing CNN predictions to the corresponding human predictions. Performance is shown as a function of $j$ in Figure \ref{fig:padding_performance_128}.\\

		From these figures, it can clearly be seen that for all types of padding, the padding-enabled-CNNs successfully learnt to make correct observations with limited data. Furthermore, when comparing the performance difference of small amounts of data with $j=2$ to full scans with $j=N$, the performance of all padding types only decreased by an average of $4\pm1\%$ and $7\pm2\%$ for mean AUROC and balanced accuracy respectively. Given that the balanced accuracy and AUROC values are significantly better than the 0.25 and 0.50 of guessing respectively, it is entirely possible to assess SPM tip state with only a small number of linescans.\\
		
		However, at $j=N$ the padding-enabled-CNNs performed significantly worse than an identical ensemble trained without padding. Here, padding reduced full size performance by up to $12\%$ and $22\%$ for the mean AUROC and balanced accuracy respectively, when compared to the worst performing padding methods. Giving CNNs the ability to classify partial scans therefore significantly harms performance, reducing the real-world effectiveness of such systems. We also note that this architecture also performed better than the ensembles presented in Gordon \emph{et al.}\cite{Gordon2019}. The large initial convolutional window may have caused this. Besides the reduced maximum performance, there was also a large computational inefficiency due to training the CNNs to perform (and subsequently ignore) a large number of expensive calculations on meaningless data. \\
		
\begin{figure}[htb!]
					\centering
					\includegraphics{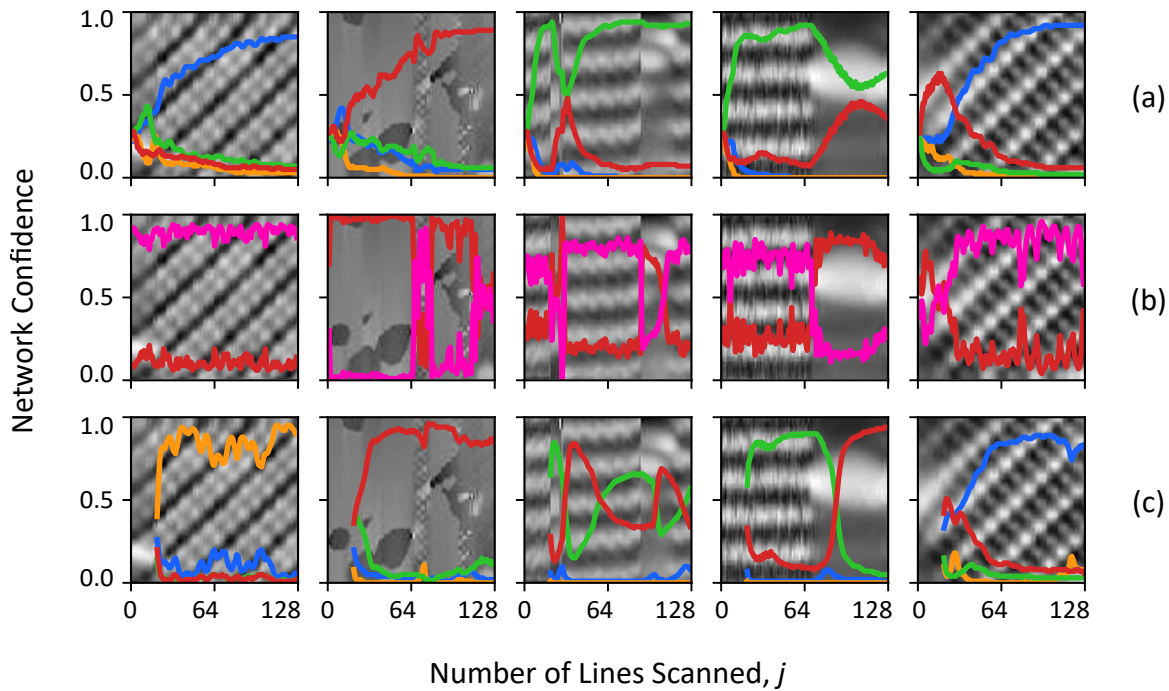}
					\caption{Figure to demonstrate a variety of methods by which the H:Si(100) tip states of individual atoms (yellow), asymmetries/dimers (blue), rows (green), and generic defects (red) can be recognised from incomplete SPM scans. Instead of detecting SPM tip states using complete scans, neural networks were taught to recognise partial scans by zero padding (a), or by classifying single linescans (b). In this case, non-defect categories had to be combined together (pink). However, optimal results were found by forming a "window" with a small group of 20 consecutive linescans, and giving additional predictive power by using a second LSTM network as the window is "rolled" over time (c). This network was found to perform the strongest at single class classifications, and showed good responsiveness with varying tip state.}
					\label{fig:sample_outputs}
				\end{figure}

		One advantage of partial-scan methods is that tip changes can be instantly detected by looking for changes and impulses in CNN output, as visible in Figure \ref{fig:sample_outputs}. This is a significant improvement on previous full-scan methods which require a secondary "tip change" network\cite{Gordon2019}. We note that without manual labelling of all tip change locations on all images, a quantitative analysis of tip-change detection is not possible. However, the imperfect ignoring of the marker values meant that some of the horizontal shears due to tip changes caused little-to-no-change in network output. The change in prediction to reflect a new tip state was also often small, and tended to "drift" rather than instantly "snap" to the new value. This was particularly problematic for tip changes later on in a scan. One explanation is that the network learnt to heavily rely on earlier scan-lines because training images often had early scanlines present, but later scanlines did so increasingly rarely. It was also impossible to detect tip changes using the "tile" method of data padding, which created a horizontal shear (visually identical to a tip change shear) between every tile. As such, padding should only be employed early on in scans and when the tip state is likely stable. 
		
		
	\subsection{Individual Linescan Windows and Cumulative Averages}
\begin{figure}[htb!]
			\centering
			\includegraphics{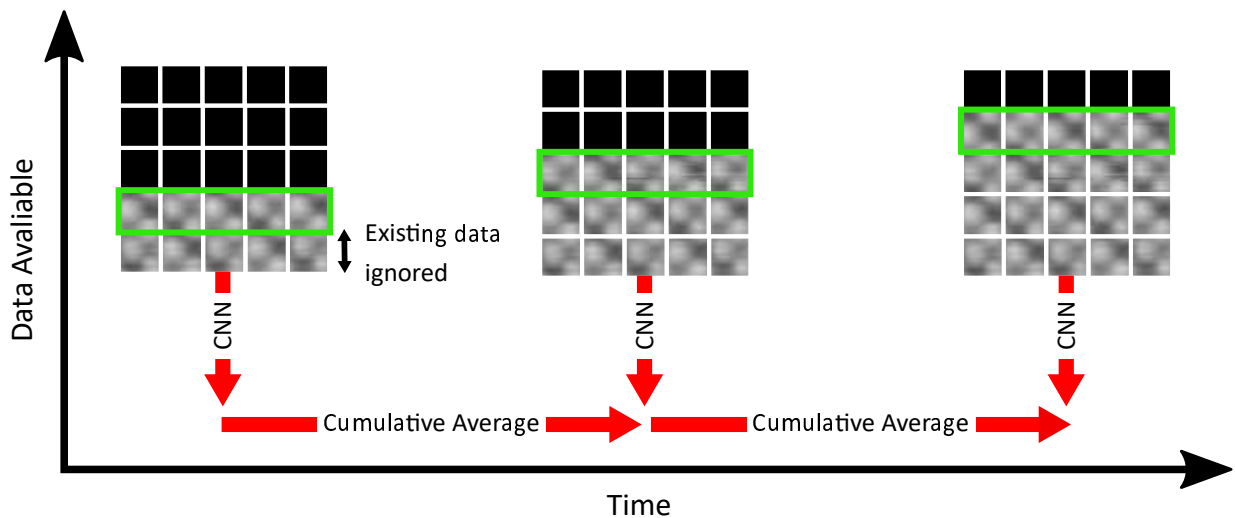}
			\caption{Figure to demonstrate a potential method to allow neural networks to predict the state of an SPM tip using incomplete scans. Instead of training/predicting with complete scans, the network (green border) was allowed to predict individual linescans. As more linescans become available during a scan, network predictions are cumulatively averaged to give context between successive linescans.}
			\label{fig:linescans_method}
		\end{figure}

		One alternative to padding incomplete scans is to train so as to classify the individual linescans that form an image, rather an image in its entirety. As new lines are scanned, they could immediately be predicted. This negates much of the insensitivity and computational wastefulness caused as a result of padding, and is demonstrated in Figure \ref{fig:linescans_method}.\\
		
		However, one consequence of basing predictions on individual linescans is that each linescan is stripped of its context to the rest of the scan. Acquiring more linescans should therefore not improve network performance. As such, a small amount of context can be applied to the other scanlines in the image by applying an additional layer to cumulatively average the network predictions using the equation
		\begin{equation}\label{eq:cum_average_linescan}
			\mathbf{P}(\mathbf{A}^i_j) = \frac{\sum_{k=1}^{j}\mathbf{P}(\mathbf{A}^i_k)}{j},
		\end{equation}
		where $\mathbf{P}(\mathbf{A}^i_j)$ is the cumulatively averaged vector describing the predictions of the $j^{th}$ linescan of the $i^{th}$ image in the dataset. A prediction for the entire image is therefore found when the condition $j=N$ is met. \\
		
		We also note that although the cumulative averaging provided context to the predictions, the actual predictive part of the network was unaware of the surrounding linescans. Whilst this averaging therefore served to reward consistent single-class output, it should be expected to have poor responsiveness to scans where the tip constantly changes shape. Furthermore, the network had little-to-no ability to distinguish between features that cannot be distinguished at the 1D level. For example, a single linescan of 'atoms' or 'rows' features in Figure \ref{fig:tip_states} would appear identical with a half-rectified sinusoid. The varying scan areas of the dataset required to make predictions invariant to scan area then prevent the network from learning any spatial information to distinguish between the two states. As such, the number of tip states was simplified to just two - "generic defect" and "visible resolution".\\
		
		Adaptions also had to be made to the network architecture. Because 2D convolutions cannot be performed on 1D data, the 2D layers of the network were replaced with their one-dimensional counterparts to provide the closest possible comparison between the protocols. Furthermore, because successive lines were often highly similar, only 1 in every 30 lines of each image were used during training to prevent improper training and decrease training time.\\
		
\begin{figure}[htb!]
			\centering
			\includegraphics[width=.5\linewidth]{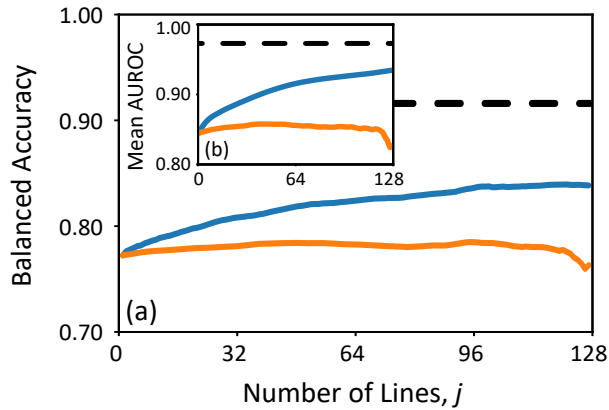}
			\caption{Figure to demonstrate the balanced accuracy (a), and mean area-under the receiver-operator-characteristic curve (b) of a neural network trained to classify the SPM tip states of the H:Si(100) surface with incomplete scans. Given that SPM data is generated one line at a time, the identical network can be trained on single linescans, instead of only on complete scans (black). In this way, the data requirements for neural network automated state detection can be reduced significantly. Because this method removes context between scans, predictions can not be influenced by prior scans. Performance is therefore only improved with the addition of new data by cumulatively averaging successive linescans in an image together (blue). Without this averaging step (yellow), performance remains roughly consistent, as expected.}
			\label{fig:linescans_performance_128}
		\end{figure}

		As before, performance was verified by iteratively predicting additional lines of the $i$ images in the holdout set and calculating the cumulative predictions using Equation \ref{eq:cum_average_linescan}. This is shown in Figure \ref{fig:linescans_performance_128}. To compare with full-sized performance, the 1D convolutions were replaced with their 2D equivalents (as used in Section \ref{sect:pad_mask}), and trained to recognise only the two simplified states.\\
		
		Without the cumulative averaging layer, the low standard deviation demonstrated that performance remained near constant as expected, with AUROC of $0.853 \pm 0.006$ and balanced accuracy of $0.780 \pm 0.004$. Un-averaged individual linescans therefore provide an effective means of making a basic, but accurate, assessment of the tip. Further, despite forcing the simplification of classes recognised, the decoupling of the lines meant that the network was highly sensitive to tip changes. This is visible when looking at the unaveraged output in Figure \ref{fig:sample_outputs}(b). As this network was clearly more responsive to state changes than padding, it is possible to use the single linescan network, (along with its low computational cost), solely for the purpose of detecting tip changes by looking for sharp peaks and changes in network output.\\
		
		Regardless, even stronger performance was seen with single-class images after cumulatively averaging. After including the layer, performance began to improve as expected, with AUROC substantially improving by 13.2\% relative to the average, and the balanced accuracy increasing by 9.9\%. This resulted in an AUROC of over 0.9, thus demonstrating highly effective ability when full data is available. This was also found to hold true for the padding strategy with all 128 linescans. It should, however, be stressed, that relative to training only with complete scans, peak performance is still reduced. In this case, when training the 2D CNN solely with complete scans and the two simplified categories, AUROC performance was near perfect, at 0.973. Further, cumulative averaging caused predictions to be significantly less sensitive to tip changes, as expected.\\
		

	\subsection{Multiple Linescan Windows and LSTM}
		Whilst single linescans provide an effective method to make a basic assessment of the tip, the inability to assess the complete range of states makes it of limited use. To overcome the lack of context between linescans, a CNN could instead be trained to recognise a small "window" consisting of a fixed number, $W$, of linescans. As new data becomes available, the window could then be "rolled" to consist of the new line and the $(W-1)$ linescans preceding it. This window could then be iteratively rolled while an image is being generated. We therefore modify Equation \ref{eq:cum_average_linescan}, and use cumulative averaging to make predictions after each successive linescan from $j=W+1$ to $j=N$ 
		\begin{equation}\label{eq:cum_average_cnnscan}
			\mathbf{P}(\mathbf{A}^i_j) = \frac{\sum_{k={W+1}}^{j}\mathbf{P}(\mathbf{A}^i_{(k-W) : k})}{j}.
		\end{equation}
		
\begin{figure}[htb!]
			\centering
			\includegraphics{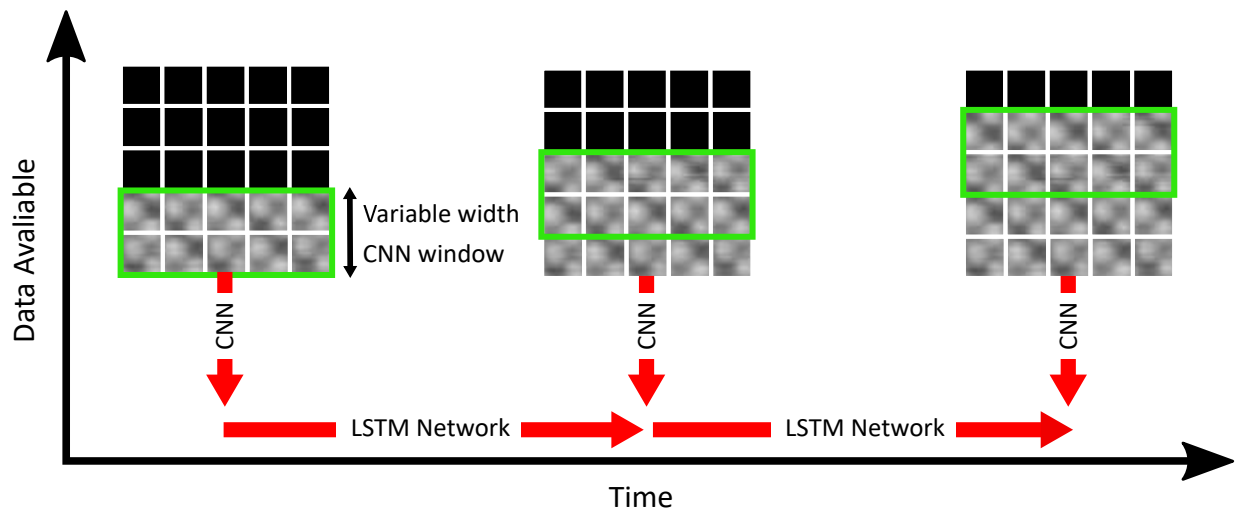}
			\caption{Figure to demonstrate a potential method to allow neural networks to predict the state of an SPM tip using incomplete scans. Rather than training/predicting with complete scans, the network (green border) can instead be allowed to predict a small group of individual linescans. This window of CNNs can then be rolled to make additional predictions as successive linescans become available over time. The outputs of these CNNs can then be fed into a second temporal neural network, to make a final prediction.}
			\label{fig:rolling_window}
		\end{figure}

		However, whilst effective at improving single-state classification performance and rewarding tip consistency, cumulative averaging does not make the predictive part of the neural network aware of the lines surrounding each window, resulting in decreased responsiveness. One recent advance in the area of video content recognition is the Long-Term Recurrent Convolutional Network (LRCN)\cite{Donahue2015}, which has been shown to be highly effective at this task. Here, a second network is placed just before the final (dense) CNN layer (which reduces the output to a size equal to the number of classification categories). This second network is typically a long-short-term-memory (LSTM) network\cite{Hochreiter1997}, which is often used for 1D sequence classification. The LSTM network then acts on the temporal domain of the data, giving context to the single CNNs which have no knowledge of how the video frames link together. This can be made analogous to SPM, where each sub-image of width $W$ becomes a video frame. The temporal element is seen as the window rolls when $j$ increments over time with new data. We therefore replace the cumulative averaging layer with an LSTM network with 256 hidden layers, and calculate predictions, $\mathbf{P}(\mathbf{A}^i_{(j-W):j})$, from $j=W+1$ to $j=N$ as before, with increasing $j$ chosen as the temporal axis. For consistency, we employ the same 2D CNN architecture as before. The resulting protocol is shown in Figure \ref{fig:rolling_window}. \\
				
		One consequence of this method is that $W$ linescans must first be accumulated before any predictions can be made. As such, whilst larger $W$ will give the network more data with which to make predictions, a larger number of linescans are required to be scanned before the window can be fully filled. For example, for a window of $W=20$, predictions can only be made after the 20th, 21st, 22nd linescans, and so on. We also note that the size and number of convolutions used meant that predictions with $W<20$ were not possible with the CNN structure used. Furthermore, all but one linescan of data is repeated with each additional window, multiplying memory usage by $N-W+1$ times.\\
		
\begin{figure}[htb!]
			\centering
			\includegraphics[width=.5\linewidth]{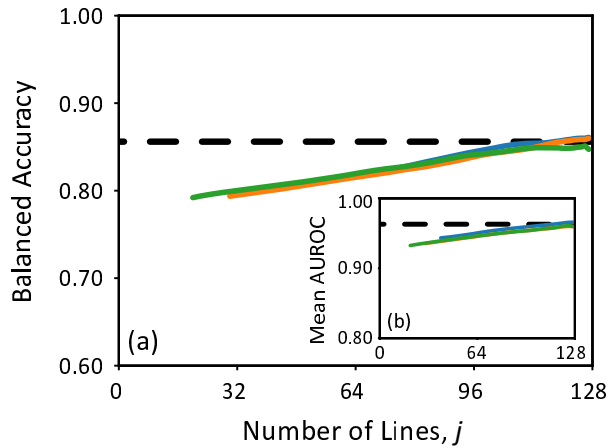}
			\caption{Figure to demonstrate the balanced accuracy (a), and mean area-under the receiver-operator-characteristic curve (b) of a neural network trained to classify the SPM tip states of the H:Si(100) surface with incomplete scans. Given that SPM data is generated one line at a time, the identical network can be trained with small groups of 20 (green), 30 (yellow), or 40 (blue) linescans, for example. These predictions are then fed into a secondary LSTM network that acts temporally. This prevents the need to train (and therefore classify) only on complete scans (black). In this way, the data requirements for neural network automated state detection can be reduced significantly.}
			\label{fig:multilinescans_performance_128}
		\end{figure}

		As can be seen in Figure \ref{fig:multilinescans_performance_128}, the inclusion of additional linescans once again resulted in improved performance, demonstrating that the LSTM component did indeed learn from the temporal evolution of the scans. Performance was also very strong regardless of $j$. For example, full scans with $W=20$ yielded a near-perfect AUROC of 0.960 and a balanced accuracy of 0.847. This is almost identical to the AUROC and balanced accuracy of 0.963 and 0.856 respectively calculated when training the CNN component only on full-sized images. The wider LRCN networks were even able to exceed full-size performance, \textit{despite using less data}. This is understandable, given that a human operator will often look not only at the scanlines, but also at how they evolve over time. Only the LRCN network takes advantage of this temporal context. It can therefore be concluded that by adding LSTM to an existing network and retraining on partial scans of fixed size, a full set of STM image classes/tip states can be correctly and accurately assessed with negligible performance impact despite using a fraction of the data. However, increasing $W$ beyond $W=30$ did not always improve performance. Although wider windows provided more opportunities to observe trends in the 2D convolutional domain, smaller windows provided more temporal elements for the LSTM layer to use.\\
		
		The benefit of using temporal information can also be seen by comparing LRCN to cumulative averaging.	For the same $W=20$ window, full-scan performance using cumulative averaging was calculated to have AUROC of 0.880 and balanced accuracy of 0.620. Not only was this slightly worse than the padding method of Figure \ref{fig:padding_performance_128}, but also significantly poorer than LRCN, which scored 9.10\% higher for AUROC and 36.7\% for balanced accuracy. This performance disparity held true regardless of values of $W$ and $j$, or when classifying variable state images. As seen in Figure \ref{fig:sample_outputs}(c), cumulative averaging was often unresponsive to both sudden changes in state. Moreover, LRCN was more able to correctly distinguish between atoms and asymmetries, and was less likely to mistakenly see rotated surfaces as a "generic defect" compared to the baseline of full scan classification. Whilst it would seem obvious to combine both LRCN and cumulative averaging, the issues with decreased responsiveness later in a scan remain. This resulted in a small performance penalty which increased as more linescans were simulated (on the order of 1\% at $j=N$). Whilst cumulative averaging was still better than guessing and is therefore another potential method for speeding up tip state recognition, LRCN is superior. \\
		
		Furthermore, whilst the state of the tip was still successfully observed with $W=20$, the size and number of convolutions used meant that window sizes below $W=20$ were not possible to test. This meant that $j=20$ lines must first be acquired before predictions can be made. To reduce the number of linescans further, larger images could instead be considered (which in this case would be achieved by downscaling from 512x512 to a size larger than 128x128). For example, simulating $W=20$ with 256 points per linescan would be equivalent to 128 points per linescan with $W=10$. However, the same number of data-points would need to be acquired before predictions could be made. There would therefore be no improvement to tip assessment speed in practise. Although the network parameters could be decreased to allow for smaller $W$, this would result in a different network that could not be fairly compared in this study. To allow for predictions at any $j$, it is trivial to create a "hybrid" network ensemble in which a basic assessment is made using the linescan/padding methods for low $j$, and then LRCN for the remainder of the scan.
	
\section{Conclusion}

	By comparing a variety of methods based around a common VGG network, we have successfully demonstrated that STM images of the H:Si(100) surface can be accurately assessed using partial scans. As such, only a few lines from a typical 128x128 scan are now required to assess the tip, which is a fraction of the data required by previous CNN assessment protocols. Given that the majority of the time spent maintaining SPM tips is spent acquiring data, a "hybrid" approach combining individual linescans and LRCN prediction would speed up CNN routines by approximately 100 times. This allows for state recognition in a time similar to that of current manual means, thus making it practical for everyday use. However, given that the states considered only apply to the H:Si(100) surface, new datasets and networks must be manually created and trained for each surface, making this strategy non-applicable to poorly understood surfaces. \\
	
	Relative to a full-size network, we find that similar or better performance can be achieved with less data by creating a small window of multiple linescans, and adding an LSTM layer to make predictions as the window is rolled over time. Furthermore, we qualitatively demonstrate that the use of partial linescans allows tip changes to be detected without the need for a secondary network. We also show that this method allows for the detection of images in which tip changes cause multiple tip states to be present, alongside their relative position in the image.	However, the low number of human classifiers and lack of manual labelling of these positions during data collection meant that only single tip-state images were quantitatively assessed. Furthermore, none of these approaches overcome the limitation of only being able to automate assessment of a single, already known surface reconstruction after a lengthy data collection process.\\
	
	In future, we aim to assess SPM tips with a "hybrid" approach combining multiple protocols of predicting with padded full-scans, individual linescans, and temporally connected partial scans of fixed width. Ultimately, this will enable seamless, automatic and constant maintenance of SPM tip integrity as part of routine experimental sessions. Unsupervised learning is the next, obvious, protocol to adopt in order to make machine learning strategies sample-independent. 

\section*{Acknowledgements}
	The authors gratefully acknowledge funding by the Engineering and Physical Sciences Research Council via grant EP/N02379X/1. We also thank I Swart, L. Knijff, S.E. Freeney, and S. Zevenhuizen of the Debye Institute for Nanomaterials Science, at Utrecht University for their continued assistance and advice (including support for the invaluable MATE-for-Dummies and access2TheMatrix Python packages.) We gratefully acknowledge helpful discussions with Bob Wolkow, John Randall, Morten Moller (who also provided the dataset of H:Si(100) images used in this work), and Richard Woolley.

\newpage
\section*{References}
\bibliographystyle{ieeetr}
\bibliography{references}

\begin{thebibliography}{10}

\bibitem{Tajaddodianfar2018}
F.~Tajaddodianfar, S.~O.~R. Moheimani, J.~Owen, and J.~N. Randall, ``{On the
  effect of local barrier height in scanning tunneling microscopy: Measurement
  methods and control implications},'' {\em {Rev. Sci. Instr.}}, vol.~{89},
  {JAN} {2018}.

\bibitem{Tewari2017}
S.~Tewari, K.~M. Bastiaans, M.~P. Allan, and J.~M. van Ruitenbeek, ``{Robust
  procedure for creating and characterizing the atomic structure of scanning
  tunneling microscope tips},'' {\em {Beilstein J. Nanotech.}}, vol.~{8},
  pp.~{2389--2395}, {NOV 13} {2017}.

\bibitem{Giessibl2019}
F.~J. Giessibl, ``The qplus sensor, a powerful core for the atomic force
  microscope,'' {\em Review of Scientific Instruments}, vol.~90, no.~1,
  p.~011101, 2019.

\bibitem{Gross2009}
L.~Gross, F.~Mohn, N.~Moll, P.~Liljeroth, and G.~Meyer, ``The chemical
  structure of a molecule resolved by atomic force microscopy,'' {\em Science},
  vol.~325, no.~5944, pp.~1110--1114, 2009.

\bibitem{Sun2011}
Z.~Sun, M.~P. Boneschanscher, I.~Swart, D.~Vanmaekelbergh, and P.~Liljeroth,
  ``{Quantitative Atomic Force Microscopy with Carbon Monoxide Terminated
  Tips},'' {\em {Phys. Rev. Lett.}}, vol.~{106}, {JAN 27} {2011}.

\bibitem{Chiutu2012}
C.~Chiutu, A.~M. Sweetman, A.~J. Lakin, A.~Stannard, S.~Jarvis, L.~Kantorovich,
  J.~L. Dunn, and P.~Moriarty, ``{Precise Orientation of a Single C-60 Molecule
  on the Tip of a Scanning Probe Microscope},'' {\em {Phys. Rev. Lett.}},
  vol.~{108}, {JUN 26} {2012}.

\bibitem{Meyer2001}
G.~Meyer, L.~Bartels, and K.~Rieder, ``{Atom manipulation with the STM:
  nanostructuring, tip functionalization, and femtochemistry},'' {\em {Comp.
  Mat. Sci.}}, vol.~{20}, pp.~{443--450}, {MAR} {2001}.

\bibitem{Gross2011}
L.~Gross, N.~Moll, F.~Mohn, A.~Curioni, G.~Meyer, F.~Hanke, and M.~Persson,
  ``{High-Resolution Molecular Orbital Imaging Using a p-Wave STM Tip},'' {\em
  {Phys. Rev. Lett.}}, vol.~{107}, {AUG 15} {2011}.

\bibitem{Jarvis2012}
S.~Jarvis, A.~Sweetman, J.~Bamidele, L.~Kantorovich, and P.~Moriarty, ``{Role
  of orbital overlap in atomic manipulation},'' {\em {Phys. Rev. B}},
  vol.~{85}, {JUN 7} {2012}.

\bibitem{Rashidi2018}
M.~Rashidi and R.~A. Wolkow, ``Autonomous scanning probe microscopy in situ tip
  conditioning through machine learning,'' {\em ACS Nano}, 2018.

\bibitem{Rashidi2019}
M.~Rashidi, J.~Croshaw, K.~Mastel, M.~Tamura, H.~Hosseinzadeh, and R.~A.
  Wolkow, ``Autonomous atomic scale manufacturing through machine learning,''
  {\em arXiv preprint arXiv:1902.08818}, 2019.

\bibitem{Gordon2019}
O.~Gordon, P.~D'Hondt, L.~Knijff, S.~Freeney, F.~Junqueira, P.~Moriarty, and
  I.~Swart, ``{Scanning Probe State Recognition With Multi-Class Neural Network
  Ensembles},'' {\em arXiv preprint arXiv:1903.09101}, 2019.

\bibitem{Zhang2019}
Y.~Zhang, A.~Mesaros, K.~Fujita, S.~D. Edkins, M.~H. Hamidian, K.~Ch'ng,
  H.~Eisaki, S.~Uchida, J.~C.~S. Davis, E.~Khatami, and E.-A. Kim, ``{Machine
  learning in electronic-quantum-matter imaging experiments},'' {\em {Nature}},
  vol.~{570}, pp.~{484+}, {JUN 27} {2019}.

\bibitem{Straton2014}
J.~C. Straton, T.~T. Bilyeu, B.~Moon, and P.~Moeck, ``Double-tip effects on
  scanning tunneling microscopy imaging of 2d periodic objects: unambiguous
  detection and limits of their removal by crystallographic averaging in the
  spatial frequency domain,'' {\em Crystal Research and Technology}, vol.~49,
  no.~9, pp.~663--680, 2014.

\bibitem{Woolley2011}
R.~A.~J. Woolley, J.~Stirling, A.~Radocea, N.~Krasnogor, and P.~Moriarty,
  ``Automated probe microscopy via evolutionary optimization at the atomic
  scale,'' {\em Applied Physics Letters}, vol.~98, p.~253104, jun 2011.

\bibitem{Stirling2013}
J.~Stirling, R.~A. Woolley, and P.~Moriarty, ``Scanning probe image wizard: A
  toolbox for automated scanning probe microscopy data analysis,'' {\em Review
  of Scientific Instruments}, vol.~84, no.~11, p.~113701, 2013.

\bibitem{Wang2016}
Y.~Wang, J.~I. Kilpatrick, S.~P. Jarvis, F.~M.~F. Boland, A.~Kokaram, and
  D.~Corrigan, ``Double-tip artifact removal from atomic force microscopy
  images,'' {\em IEEE Transactions on Image Processing}, vol.~25,
  pp.~2774--2788, June 2016.

\bibitem{Wolkow1992}
R.~A. Wolkow, ``Direct observation of an increase in buckled dimers on si (001)
  at low temperature,'' {\em Physical review letters}, vol.~68, no.~17,
  p.~2636, 1992.

\bibitem{Sweetman}
A.~Sweetman, J.~Stirling, S.~P. Jarvis, P.~Rahe, and P.~Moriarty, ``Measuring
  the reactivity of a silicon-terminated probe,'' {\em Physical Review B},
  vol.~94, sep 2016.

\bibitem{Sweetman2012}
A.~Sweetman, S.~Jarvis, R.~Danza, and P.~Moriarty, ``Effect of the tip state
  during qplus noncontact atomic force microscopy of si (100) at 5 k: Probing
  the probe,'' {\em Beilstein journal of nanotechnology}, vol.~3, p.~25, 2012.

\bibitem{Moller2017}
M.~M{\o}ller, S.~P. Jarvis, L.~Gu{\'{e}}rinet, P.~Sharp, R.~Woolley, P.~Rahe,
  and P.~Moriarty, ``Automated extraction of single h atoms with {STM}: tip
  state dependency,'' {\em Nanotechnology}, vol.~28, p.~075302, jan 2017.

\bibitem{WalshHersam2009}
M.~A. Walsh and M.~C. Hersam, ``{Atomic-Scale Templates Patterned by Ultrahigh
  Vacuum Scanning Tunneling Microscopy on Silicon},'' {\em {Ann. Rev. Phys.
  Chem.}}, vol.~{60}, pp.~{193--216}, {2009}.

\bibitem{Shen1995}
T.~Shen, C.~Wang, G.~Abeln, J.~Tucker, J.~Lyding, P.~Avouris, and R.~Walkup,
  ``{Atomic-scale desorption through electronic and vibrational-excitation
  mechanisms},'' {\em {Science}}, vol.~{268}, pp.~{1590--1592}, {JUN 16}
  {1995}.

\bibitem{Lopinski2000}
G.~Lopinski, D.~Wayner, and R.~Wolkow, ``Self-directed growth of molecular
  nanostructures on silicon,'' {\em Nature}, vol.~406, no.~6791, p.~48, 2000.

\bibitem{Fuechsle2012}
M.~Fuechsle, J.~A. Miwa, S.~Mahapatra, H.~Ryu, S.~Lee, O.~Warschkow, L.~C.
  Hollenberg, G.~Klimeck, and M.~Y. Simmons, ``A single-atom transistor,'' {\em
  Nature nanotechnology}, vol.~7, no.~4, p.~242, 2012.

\bibitem{Weber2012}
B.~Weber, S.~Mahapatra, H.~Ryu, S.~Lee, A.~Fuhrer, T.~C.~G. Reusch, D.~L.
  Thompson, W.~C.~T. Lee, G.~Klimeck, L.~C.~L. Hollenberg, and M.~Y. Simmons,
  ``{Ohm's Law Survives to the Atomic Scale},'' {\em {Science}}, vol.~{335},
  pp.~{64--67}, {JAN 6} {2012}.

\bibitem{Achal2017}
R.~Achal, M.~Rashidi, J.~Croshaw, D.~Churchill, M.~Taucer, T.~Huff,
  M.~Cloutier, J.~Pitters, and R.~A. Wolkow, ``{Lithography for robust and
  editable atomic-scale silicon devices and memories},'' {\em {Nature Comms.}},
  vol.~{9}, {JUL 23} {2018}.

\bibitem{Huff2018}
T.~Huff, H.~Labidi, M.~Rashidi, L.~Livadaru, T.~Dienel, R.~Achal, W.~Vine,
  J.~Pitters, and R.~A. Wolkow, ``{Binary atomic silicon logic},'' {\em {Nature
  Electronics}}, vol.~{1}, pp.~{636--643}, {DEC} {2018}.

\bibitem{goodfellow2016deep}
I.~Goodfellow, Y.~Bengio, and A.~Courville, {\em Deep learning}.
\newblock MIT press, 2016.

\bibitem{Fawcett2006}
T.~Fawcett, ``An introduction to roc analysis,'' {\em Pattern recognition
  letters}, vol.~27, no.~8, pp.~861--874, 2006.

\bibitem{scikit-learn}
F.~Pedregosa, G.~Varoquaux, A.~Gramfort, V.~Michel, B.~Thirion, O.~Grisel,
  M.~Blondel, P.~Prettenhofer, R.~Weiss, V.~Dubourg, J.~Vanderplas, A.~Passos,
  D.~Cournapeau, M.~Brucher, M.~Perrot, and E.~Duchesnay, ``Scikit-learn:
  Machine learning in {P}ython,'' {\em Journal of Machine Learning Research},
  vol.~12, pp.~2825--2830, 2011.

\bibitem{Young2018}
T.~Young, D.~Hazarika, S.~Poria, and E.~Cambria, ``Recent trends in deep
  learning based natural language processing,'' {\em ieee Computational
  intelligenCe magazine}, vol.~13, no.~3, pp.~55--75, 2018.

\bibitem{Simonyan2014}
K.~Simonyan and A.~Zisserman, ``Very deep convolutional networks for
  large-scale image recognition,'' {\em arXiv preprint arXiv:1409.1556}, 2014.

\bibitem{Donahue2015}
J.~Donahue, L.~Anne~Hendricks, S.~Guadarrama, M.~Rohrbach, S.~Venugopalan,
  K.~Saenko, and T.~Darrell, ``Long-term recurrent convolutional networks for
  visual recognition and description,'' in {\em Proceedings of the IEEE
  conference on computer vision and pattern recognition}, pp.~2625--2634, 2015.

\bibitem{Hochreiter1997}
S.~Hochreiter and J.~Schmidhuber, ``Long short-term memory,'' {\em Neural
  computation}, vol.~9, no.~8, pp.~1735--1780, 1997.

\end{thebibliography}

\end{document}